\documentclass{emulateapj}

\newcommand{\kms}{\, \rm{km}\,  \rm{s}^{-1}}

\def\ltsima{$\; \buildrel < \over \sim \;$}
\def\lsim{\lower.5ex\hbox{\ltsima}}
\def\Msunh{\mbox{$h^{-1}$M$_\odot$}}
\def\mpch{\mbox{$h^{-1}$Mpc}}

\def\Mpc{{\rm Mpc}}

\def\deg{\ifmmode{^\circ}\else{$^\circ$}\fi}
\def\hGpc{\ifmmode{h^{-1}{\rm Gpc}}\else{$h^{-1}{\rm Gpc}$}\fi}
\def\hkpc{\ifmmode{h^{-1}{\rm kpc}}\else{$h^{-1}{\rm kpc}$}\fi}
\def\hMpc{\ifmmode{h^{-1}{\rm Mpc}}\else{$h^{-1}{\rm Mpc}$}\fi}
\def\hMsun{\ifmmode{h^{-1}M_\odot}\else{$h^{-1}M_\odot$}\fi}

\def\kms{{{\rm km}/{\rm s}}}

\def\muK{\ifmmode{\mu{\rm{K}}}\else{$\mu$K}\fi}
\def\mum{\ifmmode{\mu{\rm{m}}}\else{$\mu$m}\fi}

\shorttitle{The orientation of dark matter halos}

\shortauthors{Patiri et al.}

\begin{document}

\title{The alignment of dark matter halos with the cosmic web}

\author{Santiago G. Patiri\altaffilmark{1}, Antonio J. Cuesta\altaffilmark{2},
 Francisco Prada\altaffilmark{2}, Juan Betancort-Rijo\altaffilmark{1,3} and Anatoly Klypin\altaffilmark{4}}

\altaffiltext{1}{Instituto de Astrof\'{\i}sica de Canarias, Calle V\'{\i}a
L\'actea, E-38200, La Laguna, Tenerife, Spain; spatiri@iac.es}
\altaffiltext{2}{Instituto de Astrof\'{\i}sica de Andaluc\'{\i}a (CSIC), E-18008, Granada, Spain; ajcv@iaa.es; fprada@iaa.es}
\altaffiltext{3}{Facultad de Fisica, Universidad de La Laguna, Astrof\'{\i}sico Francisco S\'anchez, s/n, La Laguna Tenerife E38200, Spain; jbetanco@iac.es}
\altaffiltext{4}{Astronomy Department, New Mexico State University, MSC 4500, P.O.
Box 30001, Las Cruces, NM 880003-8001; aklypin@nmsu.edu}

\begin{abstract}

We investigate the orientation of the axes and angular momentum of dark matter halos with 
respect to their neighboring voids using high resolution N-body cosmological simulations. 
We find that the minor axis of halos tends to be aligned along the line joining the halo with the center 
of the void, and that the major axis is perpendicular to this line. However, we find that the 
angular momentum of halos does not have any particular orientation.
These results may provide information about the mechanisms whereby the large-scale structure of the 
Universe affects galaxy formation and cast light upon the issue of the orientation of galaxy disks with 
respect to their host halos.

\end{abstract}
\keywords{galaxies: halos  --- large--scale structure of the universe -- methods:
statistical --- dark matter}

\section{Introduction}

The knowledge of the intrinsic alignments of the angular momentum of dark matter halos 
with respect to their axes and the alignments of these axes with respect to the
galaxy disks is of great importance in order to understand the processes
involved in galaxy formation. Moreover, the alignment of halo axes, angular momentum and
galaxy spins with respect to the large-scale structure may provide further clues to this end; 
for example, they could be determinant to understand the origin of
the angular momentum of galaxies (see e.g. Navarro et al. 2004). They also may help to understand several
observational results like the distribution of galactic satellites 
(e.g. Zentner et al. 2005, Azzaro et al 2005), the formation of galactic warps (e.g. Dekel \& Shlosman 1983; Toomre
1983; Bailin 2004) and the tidal streams observed for example in the Milky Way (e.g. Ibata et al. 2002). 
Furthermore, they could be a source of contamination in weak lensing studies (e.g. Heymans et al. 2006).

Dark matter halos are in general triaxial being defined by their
three semi-axes, $a \geq b \geq c$ (Dubinski \& Carlberg 1991; Warren et al. 1992; Cole \& Lacey 1996;
Jing \& Suto 2002; Allgood et al. 2006). Recently, some effort have been devoted to measure their 
intrinsic alignments with respect to the angular momentum. It has been found
that the angular momentum of dark matter halos tends to be parallel 
to their minor axis (Bailin \& Steinmetz 2005; Allgood et al 2006). 
This situation is typical for rotationally supported objects. However, as dark matter halos are 
found to be not rotationally supported, this alignment could have a non trivial origin (Allgood et al 2006).

On the other hand, the measurement of the alignments of galaxy disks with their host dark matter halos 
is both theoretical and observationally difficult. Using cosmological hydrodynamic simulations 
of disk galaxy formation, Bailin et al. (2005) found that the halo minor axis within 0.1$R_{vir}$ 
matches the disk axis. Beyond 0.1$R_{vir}$, the halo orientation is uncorrelated with the disk axis. 
Moreover, Hoekstra, Yee \& Gladders 2004, using galaxy-galaxy lensing measurements find that galaxies 
(disks plus ellipticals combined) are aligned with the major axis of their dark matter halo.
Other works show that the angular momentum of disks and dark matter are not perfectly aligned
(van den Bosch et al. 2002; Sharma \& Steinmetz 2005). Hence, the final orientation of the galactic 
disk in a cosmological dark matter halo is still an unresolved problem.

One way of gaining insight about these alignments is to study the alignments of both, halos and disks, with 
respect to the large-scale structure. The basic idea is that 
the angular momentum of halos and galaxies is generated by tidal torques
operating on the primordial material destined to form them (Peebles 1969; White 1984).
In this scenario, the primordial angular momentum of barions and dark matter are aligned and 
the final galactic angular momentum is a `relic' of the primordial one. According to this, 
the current misalignment between galactic and dark matter halo angular momentum is due to 
subsequent tidal torques affecting dark matter halos. However, this scenario could be more complex
because additionally to the primordial tidal torques operating on the material which 
eventually becomes the disk, the mass accretion by the disk is a clumpy and 
stochastic process (see Vitvitska et al. 2002).
Along this line, Navarro et al. (2004) found, using high resolution numerical simulations, that galaxy disks are
rather well aligned with the angular momentum of the surrounding matter at the epoch of turnaround.
Bailin \& Steinmetz (2005) found that the minor axes of halos of groups and clusters show a strong tendency to
lie perpendicular to the direction of filaments. However, it has been suggested that although filaments cause 
alignments of cluster-size halos, 
they have not discernible influence on smaller halos (Altay et al. 2006). On the other hand, it has recently been found using 
observations from the 2dFGRS and the SDSS, that galactic disk planes tend to be perpendicular to the walls 
of cosmic voids (Trujillo, Carretero \& Patiri 2006).

In this letter we study, using high resolution N-body simulations, the orientation of 
axes and angular momentum of galactic size dark matter halos embedded in the walls of 
large voids, those with radius larger than $10 \mpch$. We will discuss the 
observational implications of our results. 


\section{Numerical simulations and Methodology}

\subsection{N-body simulations}

We have used for our analysis a high resolution N-body numerical 
simulation of $512^{3}$ particles in a box of $120\mpch$ on a side assuming the
concordance $\Lambda$CDM cosmological model ($\Omega_0$ = 0.3,
$\Omega_\Lambda$ = 0.7 and $H_{0} = 100h \kms/\Mpc$ with $h = 0.7$). The 
simulation was done with the Adaptive Refinement Tree code (ART, Kravtsov et al. 1997). 
The force resolution was $1.8 \hkpc$ comoving and the mass of a particle $1.1 \times 10^{9} \mpch$.
Dark matter halos are identified in the simulation by the Bound Density Maxima 
algorithm (BDM, Klypin \& Holtzman 1997; Klypin et al. 1999). For more details about this 
simulation see Table 1 ($L120_{0.9}$) in Allgood et al. (2006). There is a total of 181000 halos 
in the whole simulation box at $z=0$. However, we limited our analysis to well-resolved halos
with more than 1000 particles inside $R_{vir}$ to be able to analyze accurately
their internal axes and angular momentum orientations.

\subsection{Searching for voids and halo sample}

In order to study the alignments of dark matter halos with their 
surrounding large-scale structure, we searched for large voids 
in the numerical simulation box using the HB void finder (Patiri et al. 2006).
The HB Void Finder is conceptually simple, as it searches
for the maximal spheres with radius larger
than a given value that are empty of halos with mass larger than
some specified value. This code is designed to be accurate
and computationally efficient in finding the largest
voids in any {\it point} distributions like galaxy samples or 
dark matter halos samples. To search for these voids, the code first
generates a sphere of fixed radius randomly located within
the sample. We check if this sphere is empty of
previous selected halos. If so, we increase the size of the sphere until
its surface reaches the four nearest halos. This is our void. For a
full description of the algorithm see Patiri et al. (2006).

In the present work, we have searched for voids defined by dark matter 
halos with mass larger than $1.0 \times 10^{12} \Msunh$. We choose this mass 
in order to compare with the observational results found by Trujillo, Carretero \& Patiri (2006),
i.e. we select the halo mass so that their number density is equal to the number density of 
galaxies found in the observed galaxy samples (galaxies brighter than $M_{b_{J}}=-19.4 +5log h$). 
With these definitions, we found a total of 60 voids with radius larger than $10\mpch$ in the whole simulation box.
Once we have obtained the voids, we selected the halos located within shells of 
$4 \mpch$ thick beyond the surface of the void. Also, as we are interested in the orientations 
of galactic size halos with respect to their neighbouring large-scale structure, we restricted the mass of the halos 
to be smaller than $1.0 \times 10^{13} \Msunh$. We detect a total of 1729
halos in this mass range that are located in the shells of large voids.
For each of these halos we obtain their principal axes ($\hat{a},\hat{b},\hat{c}$) and calculate their 
angular momentum, in order to measure their orientations with respect to the void centers (see next section).

\subsection{Measuring halo orientations}

We follow the method presented in Allgood et al.(2006) to fit our halos to ellipsoids. First of all, we 
calculate the \emph{reduced} inertial tensor for all the particles inside 
$R_{\mathrm{vir}}$ as $I_{ij}=\Sigma x_ix_j/r^2$, where the sum is over every particle inside a 
sphere of radius $R_{\mathrm{vir}}$ around the center of the halo. The halo center and 
$R_{\mathrm{vir}}$ are given by our BDM halo finder. The coordinates 
here are Cartesian and the denominator (squared distance to halo center) makes the eigenvalues 
of the tensor, $\lambda_1>\lambda_2>\lambda_3$ dimensionless and non-sensitive to particle distances. 
The axis lengths $a>b>c$ are proportional to the square root of the eigenvalues. After this first step, 
we iterate this procedure to get more and more approximate ellipsoids that fit our dark matter 
distribution. To achieve it, we take the eigenvalues that we have just calculated and we normalize them so 
that $a=R_{\mathrm{vir}}$ and the ratios $c/a$, $b/a$ are preserved. We then take into account every 
particle inside the ellipsoid (the ones that obey 
$\left(\frac{\vec{r}\cdot \hat{a}}{a}\right)^2 +\left(\frac{\vec{r}\cdot \hat{b}}{b}\right)^2 +\left(\frac{\vec{r}\cdot \hat{c}}{c}\right)^2 <1$, 
with $\hat{a},\hat{b},\hat{c}$ being the unit eigenvectors), and calculate the reduced inertial tensor, defined now as:
\begin{equation}
I_{ij}=\sum \frac{x_ix_j}{r_e^2} \qquad r_e=a\sqrt{x^2/a^2+y^2/b^2+z^2/c^2}
\end{equation}

Its diagonalization results in the new eigenvalues and axis lengths. This method is repeated until 
convergence is achieved in every axis, with a tolerance $\epsilon=5 \times 10^{-2}$. We stop 
the fit after just 7 iterations, as the convergence is usually very fast. The halos that could not be fitted to ellipsoids ($\sim 8$\%) are therefore
rejected.

Once the ellipsoid axes are measured, we calculate in a straightforward way the angle between the 
vector joining the center of the halo to the center of the void 
$\vec{R}=R\hat{R}\equiv\vec{R}_{\mathrm{halo}}-\vec{R}_{\mathrm{void}}$, and the principal halo axes with 
$\alpha=\arccos \vert\hat{R}\cdot\hat{a}\vert \quad \beta=\arccos \vert\hat{R}\cdot\hat{b}\vert \quad \gamma=\arccos \vert\hat{R}\cdot\hat{c}\vert$. 
The  $1 \sigma$ error for the direction of the axes is about $8^{\circ}$ (see equation $5$ in Bailin \& Steinmetz 2004).

As the particles inside the ellipsoid are known after the iterative method converges, we can calculate 
the total angular momentum vector of the halo as:
\begin{equation}
\vec{L}=\sum m\vec{r}\times\vec{v}
\end{equation}

where $\vec{r}$ is again the particle position referred to the center of the halo, and $\vec{v}$ is 
the particle velocity. Note that we are approximating this center with the real center of mass of 
the ellipsoid, otherwise there would be non-negligible additional terms to take into account. 
Then the angle $\theta$ between $\vec{L}$ and $\vec{R}$ is calculated as follows:
\begin{equation}
\theta=\arccos \big|{\frac{\vec{R}\cdot\vec{L}}{RL}}\big|
\end{equation}

There are several sources of noise when calculating these angles, so it is not wise to take every 
halo we have in our sample. In particular, we reject 
the halos whose axes are determined to within an uncertainty greater than $0.2~ \mathrm{rad}\simeq 11^{\circ}$. 
Moreover, we reject halos with degenerate axes, as the 
direction of these axes is ill-defined and arbitrary within a plane (see Jing \& Suto 2002). Therefore, we just 
take halos with axis ratios lower than 0.95, in order to avoid such degeneracy. After all, 
we have 1187 halos left for our analysis.

We also checked if our halos have the expected properties reported in the literature for their axis ratios and 
the intrinsic alignment of the angular momentum with the direction of principal axes. 
No discrepancies were found. The mean axis ratios in our mass range are 
$c/a=0.64\pm 0.14$, $b/a=0.78\pm 0.15$, and $c/b=0.82\pm 0.09$ in agreement with Jing \& Suto (2002), 
Bailin \& Steinmetz (2005), and Allgood et al. (2006). The median direction cosines of the angular 
momentum with the principal axes are $L\cdot a=0.32$, $L\cdot b=0.48$, and $L\cdot c=0.69$, in agreement 
with the results in Bailin \& Steinmetz (2005). About 75 \% of halos have an angle between their angular 
momentum and their major axes greater than $60^{\circ}$ as shown in Gottl\"ober \& Turchaninov (2006).

\section{Results and Conclusions}

In Figure 1 we show the orientations of the three halo axes and angular momentum 
with the cosmic voids. In the top-left panel we show the probability distribution 
of the $\alpha$ angle which is defined between the major axis of the dark matter halos 
and the vector pointing to the center of the void (black dots). Similarly, in the top-right and 
bottom-left panels we present the angles $\beta$ and $\gamma$ which are the ones 
between the middle and minor axis with the direction to the center of the void 
respectively. Finally, in the bottom-right panel we show the angle $\theta$ between 
the halo angular momentum and the direction to the center of the void. The solid line 
corresponds to the probability distribution of randomly distributed angles, which is a sine function for geometrical reasons.

In these plots we clearly see that halos are not randomly oriented within the shells of voids. 
The major axis of halos tends to lie along the plane perpendicular to the radial direction, 
i.e. there is an excess of halos with $\alpha$ angles near 90 degrees. On the contrary, 
the minor axis tend to point to the direction of the center of the void. There is a 
marginal orientation of the middle axes toward the plane perpendicular to the radial direction. 
However, we do not see any particular orientation of the halo angular momentum with respect to the
large-scale structure, confirming recent results obtained also from N-Body simulations using similar 
methodology (Heymans et al. 2006).  

To test the robustness of our results we have conducted several tests. First of 
all, we have analyzed the orientations of dark matter matter halos selected as 
described above, but in this case we used ``fake'' voids, i.e. we randomly 
placed spheres of fixed radius over the sample and calculated the angles between the 
axis (and angular momentum) of the dark matter halos located in the shells of these ``fake'' voids 
and the direction to their center. As expected, we found that the angles of the halos chosen 
to be in the shells of random spheres are uniformly distributed, i.e. they follow a sine distribution.
Moreover, we applied two different statistical tests: the average of the cosine test and a standard 
Kolmogorov-Smirnov test. If the angles are randomly distributed, the average of the cosine of these angles 
is $0.5$. So that, the bigger is the deviation from this value the larger will be the effect.
In Table 1 we show the results of the statistical tests. In the upper part of the table we show the results 
for the mean cosine (column 1) and the probabilities that the results be compatible with isotropy as given by 
the mean cosine and the KS tests (columns 2 and 3) for halos located in the shells of voids and, in the bottom 
part the same results but for halos inside the shells of the random spheres (i.e. the `fake' voids). From these results, 
we conclude that we can reject with high confidence (more than 7 sigmas) the null hypothesis (i.e. that the axis of dark 
matter halos are randomly distributed) for the halos in the shells of voids. As expected, the results 
of the statistical test for the halos located in the shells of the random spheres show that the 
orientation of these halos is compatible with the random distribution.

On the other hand, we used our data to gain some insight into the nature of the internal alignment of the 
minor axis with angular momentum found in previous works (e.g. Bailin \& Steinmetz 2004). We selected 
those halos where both directions form an angle smaller that $20\deg$. 
For these halos, we found that the alignment between the minor axis and the radial direction 
of the void is weaker than for randomly chosen halos, being even compatible with a random distribution.

Comparing these results with the observed tendency for the spin of spiral galaxies to point along 
the plane perpendicular to the direction of the center of the void found by Trujillo, Carretero \& Patiri (2006), three 
scenarios are possible: 1) The non-isotropy of the orientation of the spin of the galaxies is entirely 
induced by the halo alignments, 2) both alignments are two effects with common origin (namely, the 
tidal field in the shells of voids), or 3) a combination 
of the previous ones. In the first scenario the probability distribution for the angle between the major axis 
of the host halo and the spin of their galaxy is independent of the angle between the spin and the 
radial direction of the void. In this case, this probability distribution may be uniquely determined by 
comparing the distribution of the angle between the major axis and the radial direction with that of 
the angle between the spin of the galaxies and the radial direction. This would imply that the spin is very 
closely aligned with the major axis of the host halo. However, recent hydrodynamical simulations (Bailin et al. 2005) 
does not show this effect (they found that the spin of galaxies is uncorrelated with the halo). 
Therefore, the other two scenarios seem more plausible.

\acknowledgments

We thank useful discussions with Conrado Carretero, Stefan Gottl\"ober, 
Andrey Kravtsov and Brant Robertson.
SGP, AJC, FP \& JBR thank support from the Spanish MEC under grant PNAYA 2005-07789. 
AJC acknowledges the financial support by the MEC through the spanish grant FPU AP2005-1826.


\bibliography{ms.bib}

\begin{thebibliography}{27}
\expandafter\ifx\csname natexlab\endcsname\relax\def\natexlab#1{#1}\fi

\bibitem[Allgood et al.(2006)]{brandon} Allgood, B., Flores, 
R.~A., Primack, J.~R., Kravtsov, A.~V., Wechsler, R.~H., Faltenbacher, A., 
\& Bullock, J.~S.\ 2006, \mnras, 367, 1781 

\bibitem[]{} Azzaro, M., Zentner, A.R., Prada, F. \& Klypin, A. \ 2005, submitted to \apj, astro--ph/0506547

\bibitem[Bailin(2004)]{2004PhDT} Bailin, J.\ 2004, Ph.D.~Thesis 

\bibitem[Bailin \& Steinmetz(2004)]{BailinMatado2} Bailin, J., \& Steinmetz, M.\ 2004, \apj, 616, 27

\bibitem[Bailin \& Steinmetz(2005)]{BailinMatado1} Bailin, J., \& Steinmetz, M.\ 2005, \apj, 627, 647 

\bibitem[Bailin et al.(2005)]{bailinNew} Bailin, J., et al.\ 2005, \apjl, 627, L17 

\bibitem[Cole \& Lacey(1996)]{colalacey} Cole, S., \& Lacey, C.\ 1996, \mnras, 281, 716 

\bibitem[Dekel \& Shlosman(1983)]{1983IAUS..100..187D} Dekel, A., \& 
Shlosman, I.\ 1983, IAU Symp.~100: Internal Kinematics and Dynamics of 
Galaxies, 100, 187

\bibitem[Dubinski \& Carlberg(1991)]{dubi} Dubinski, J., \& Carlberg, R.~G.\ 1991, \apj, 378, 496 

\bibitem[]{Stefan} Gottl\"ober S. \& Turchaninov V. \ 2006, EAS Publications Series, 20, 25

\bibitem[]{}  Heymans, C., White, M., Heavens, A., Vale, C. \& Van Waerbeke, L. \ 2006, submitted to \mnras, astro--ph/0604001

\bibitem[Ibata et al.(2002)]{Ibata} Ibata, R.~A., Lewis, G.~F., Irwin, M.~J., \& Quinn, T.\ 2002, \mnras, 332, 915 

\bibitem[]{JS}  Jing Y. P. \& Suto Y. \ 2002, \apj, 574, 538

\bibitem[Klypin \& Holtzman(1997)]{klypin} Klypin A. \& Holtzman J., 1997, astro--ph/9712217

\bibitem[Klypin et al.(1999)]{KlypinSat1999} Klypin A., Kravtsov A.V.,
  Valenzuela O., Prada F., 1999, \apj, 522, 82

\bibitem[Kravtsov et al.(1997)]{1997ApJS..111...73K} Kravtsov, A.~V., 
Klypin, A.~A., \& Khokhlov, A.~M.\ 1997, \apjs, 111, 73 

\bibitem{nav04}  Navarro, J. F., Abadi, M. G. \& Steinmetz M., 2004, ApJ, 613, L41

\bibitem{pat06}  Patiri S. G., Betancort-Rijo, J., Prada, F., Klypin, A. \&
Gottl$\ddot{o}$ber, S., 2006, \mnras, 369, 335

\bibitem{pee69}  Peebles, P.J.E., 1969, ApJ, 155, 393

\bibitem[Sharma \& Steinmetz (2005)]{ss05} Sharma, S., \& Steinmetz, M. 2005, \apj, in press, astro--ph/0406533

\bibitem[Toomre(1983)]{1983IAUS..100..177T} Toomre, A.\ 1983, IAU 
Symp.~100: Internal Kinematics and Dynamics of Galaxies, 100, 177 

\bibitem[Trujillo et al.(2006)]{bcp} Trujillo, I., Carretero, C., \& Patiri, S.~G.\ 2006, \apjl, 640, L111 

\bibitem[van den Bosch et~al.(2002)]{vdb-etal02} van den Bosch, F.~C., Abel, T., Croft, R.~A.~C., Hernquist, L., \&
  White, S.~D.~M. 2002, \apj, 576, 21

\bibitem[Vitvitska et~al.(2002)]{vitvitska-etal02} Vitvitska, M., Klypin, A.~A., Kravtsov, A.~V., Wechsler, R.~H.,
  Primack, J.~R., \& Bullock, J.~S. 2002, \apj, 581, 799

\bibitem[Warren et~al.(1992)]{warren-etal92} Warren, M.~S., Quinn, P.~J., Salmon, J.~K., \& Zurek, W.~H. 1992, \apj,
  399, 405

\bibitem{whi84}  White, S. D. M., 1984, ApJ, 286, 38

\bibitem[Zentner et al.(2005)]{2005ApJ...629..219Z} Zentner, A.~R., 
Kravtsov, A.~V., Gnedin, O.~Y., \& Klypin, A.~A.\ 2005, \apj, 629, 219 

\end{thebibliography}

\clearpage

\begin{deluxetable}{lccc}
\tablewidth{0pt}
\tablecaption{Results of the statistical tests}
\tablehead{ \colhead{Angle} &  \colhead{$<cos>$}   & \colhead{$<cos>$ Prob.}  & \colhead{KS Prob.}} 
\startdata
\cutinhead{ Halos in walls of voids}
 $\alpha$ & $0.459$  &  $1.13 \times 10^{-8} $  & $5.4 \times 10^{-7}$ \\
 $\beta$  & $0.481$  &  $7.58 \times 10^{-3} $  & $2.0 \times 10^{-3}$  \\
 $\gamma$ & $0.556$  &  $9.19 \times 10^{-15}$  & $9.4 \times 10^{-13}$  \\
 $\theta$ & $0.509$  &  $0.19$                  & $0.12$  \\
\cutinhead{ Halos in walls of random spheres }
 $\alpha$ & $0.496$  &  $0.48 $  & $0.23$ \\
 $\beta$  & $0.509$  &  $8.90 \times 10^{-2} $  & $7.80 \times 10^{-2}$  \\
 $\gamma$ & $0.492$  &  $0.19 $  & $0.10$  \\
 $\theta$ & $0.502$  &  $0.69 $  & $0.57$  \\
\enddata 
\end{deluxetable}

\clearpage

\begin{figure*}
\includegraphics[width=170mm]{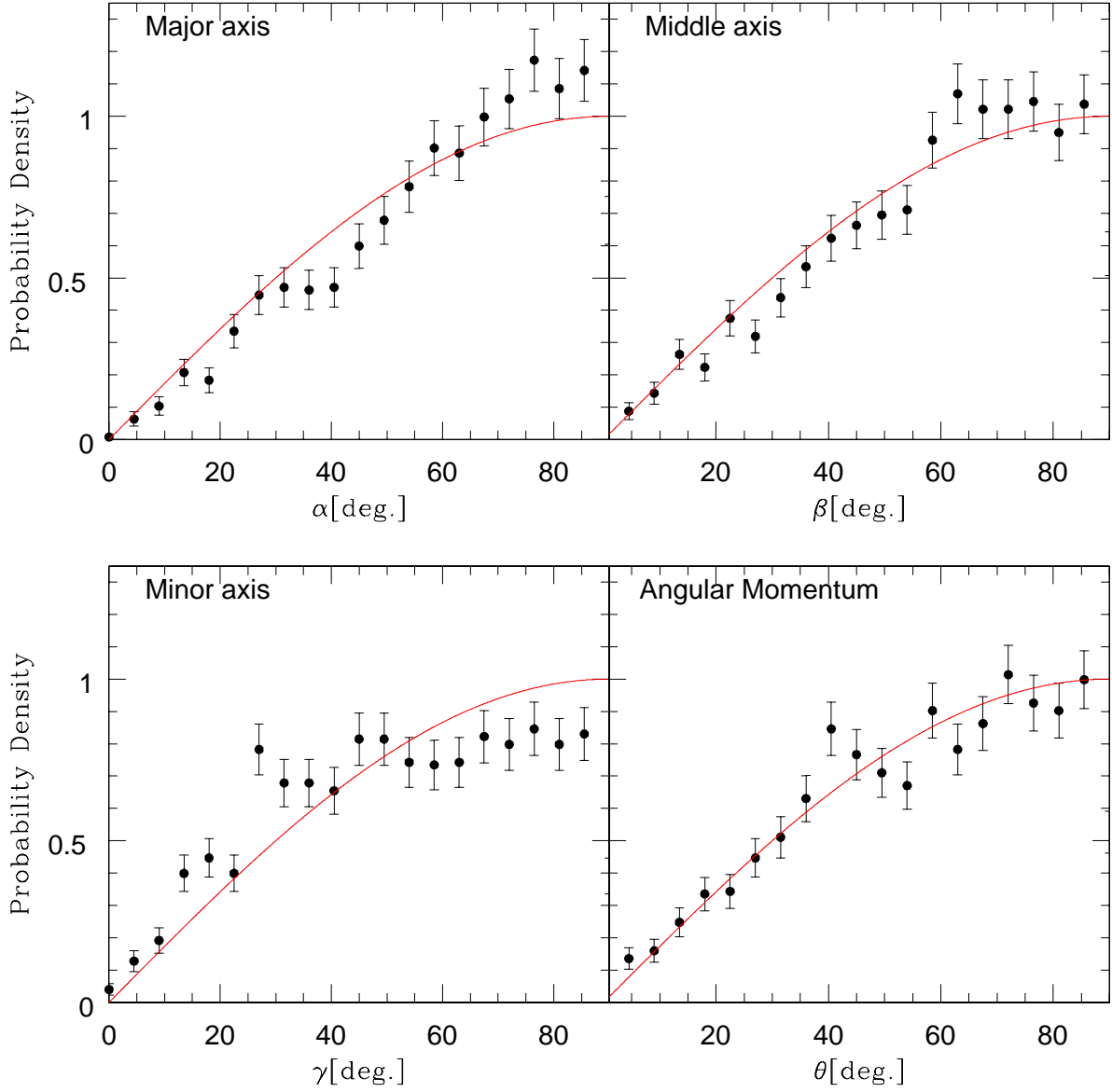}
\caption{Probability density distribution for the angles between the axes (and angular momentum) of 
the dark matter halos and the vectors pointing to the center of the voids. 
Top-left panel: $\alpha$ angle for the major axis. Top-right panel: $\beta$ angle for the middle axis. 
Bottom-left panel: $\gamma$ angle for the minor axis. Bottom-right panel: $\theta$ angle for the angular
momentum. The error bar on each bin is the Poissonian error. The solid line represents randomly distributed
angles (i.e. a sine distribution)}
\label{fig:fig1} 
\end{figure*}

\end{document}